\documentstyle[rotate,epsf]{mn}

\title[On the abundance of Lithium in T Coronae Borealis]
{On the abundance of Lithium in T Coronae Borealis}
\author[T.~Shahbaz, P.H.~Hauschildt, T.~Naylor and F.~Ringwald]{
T.~Shahbaz$^{1}$, P.H.~Hauschildt$^{2}$, T.~Naylor$^{3}$ 
and F.~Ringwald$^{4}$\\
$^{1}$University of Oxford, Department of Astrophysics, Nuclear Physics
Building, Keble Road, Oxford, OX1~3RH, UK \\
$^{2}$Department of Physics and Astronomy,
The University of Georgia, Athens, GA 30602-2451, USA \\
$^{3}$Department of Physics, Keele Universtity, Keele,
Staffordshire, ST5~5BG, UK \\
$^{4}$Deptartment of Physics and Space Sciences, Florida Institute of
Technology, 150 West University Boulevard, Melbourne, FL 32901-6988, USA }

\begin{document}

\maketitle

\begin{abstract}

\noindent
We have obtained high resolution echelle spectroscopy of the recurrent
nova T CrB. We find that the surface lithium abundance in T CrB is
signifcantly enhanced compared to field M giants, where it is not
detectable. We offer possible explanations for this in terms of either a
delay in the onset of convection in the giant star, enhanced coronal
activity due to star-spots or the enhancement of Li resulting from the
nova explosion(s).

\end{abstract}
\section{Introduction}

T Coronae Borealis (= T CrB) is a recurrent nova which underwent major
eruptions in 1866 and 1946. Its quiescent optical spectrum shows
M--type absorption features with Balmer and He emission lines, and the
Balmer jump (Kenyon 1986, and references therein). An optical spectrum
of such a type qualified T CrB to be classed as a symbiotic system.

The optical and infrared light curve show a double humped modulation
which is due to the observer seeing differing aspects of the tidally
locked, gravitationally distorted M-giant companion star. Shahbaz et
al (1997) have modelled these light variations and have concluded that
the binary system must contain an accreting white dwarf very close to
the Chandrasekhar limiting white dwarf mass of 1.4$M_{\odot}$; this is in
agreement with the high mass required to explain the outbursts in the
recurrent nova in terms of a thermonuclear runaway process (Webbink
1976; Livio \& Truran 1992).

Models of the nova explosion predict that the thermonuclear runaway
can produce large amounts of lithium (see Starrfield et al. 1978). In
this letter we present high resolution echelle spectroscopy of T
CrB. We determine the surface abundance of Li and give possible
explanations to its abundance.
 
\begin{table}
\caption{Journal of observations}
\begin{center}
\begin{tabular}{lccc}
Object   &  Date    & Exposure time      & Comments \\
         &          &                    &              \\
T CrB    &  15/6/95 &  3$\times$1800 s   &  $\phi$=0.90 \\
T CrB    &  16/6/95 &  2$\times$1800 s   &  $\phi$=0.90 \\
         &          &                    &              \\
HD~129712 &  15/6/95 &  1$\times$300 s    &  M3$\sc iii$ \\
HD~151203 &  15/6/95 &  1$\times$600 s    &  M3$\sc iii$ \\
HD~152112 &  15/6/95 &  1$\times$600 s    &  M3$\sc iii$ \\
\end{tabular}
\end{center}
\end{table}

\section{Observations and data reduction}

We obtained high resolution echelle spectra of T CrB on the nights of the
15th and 16th June 1995 using MUSICOS (MUlti-SIte COntinuous
Spectroscopy; Baudrand \& Bohm 1992) mounted on the 2-m Bernard Lyot
telescope at the Pic du Midi Observatory, France. We also observed
several template M3$\sc iii$ stars (see Table 1 for a log of the
observatons).  Finally, Th-Ar lamps and dome-flats were observed to
wavelength calibrate and flat-field the data respectively.

The spectra were extracted, wavelength and flux calibrated using the
procedures described by Baudrand \& Bohm (1992). The accuracy of the
wavelength calibration was 0.024\AA. Our observations which cover the
spectral range 5400--8800 \AA\ are composed of 43 orders, each covering
about 100 \AA\ with a velocity dispersion of 4 km~s$^{-1}$/pixel. The
velocity resolution of the spectra was 10 km~s$^{-1}$ (FWHM at
H$\alpha$). In this paper we only use the order covering the spectral
range 6680--6780 \AA. We cross correlated all the T CrB spectra in order
to determine the velocity shift relative to a standard star. The spectra
were then Doppler shifted and summed (see Figure 1).

\section{Models and Abundance Analysis}

For this analysis we use model atmospheres with a setup similar to the
``NextGen'' model grid of Hauschildt et al (1998). Our models are
spherically symmetric LTE models calculated with the general stellar
atmosphere code {\tt PHOENIX}. We use a direct opacity sampling method to
include line blanketing of both atoms and molecules. The equation of
state includes more than 500 species (atoms, ions and molecules).  For the
relatively high effective temperatures considered here, the effects of
dust condensation and opacities are negligible.  The radiative transfer
equation is solved using an operator splitting method.  Details of the
calculational methods are given in the above reference.

We used solar abundance models in the effective temperature range $3000
\leq T_{\rm eff} \leq 3600\,$ with $2.5 \leq \log(g) \leq 3.5$ as starting
point. For the best fitting model, $\log(g) = 2.5$ and $T_{\rm
eff}=3200\,$K, we calculated a set of synthetic spectra with lithium
abundances $0.0 \leq
\log_{10}\epsilon(\rm Li)\leq 3.31$. Here, $\epsilon(\rm Li)$ is the
number of lithium nuclei for each $10^{12}$ hydrogen nuclei.
$\log_{10}\epsilon(\rm Li)=1.16$ is the abundance of lithium in the solar
atmosphere and $\log_{10}\epsilon(\rm Li)=3.31$ is the meteoritic lithium
abundance. The spectra were calculated with the same code and general
setup (including spherical symmetry) that was used to calculate the
models in the model atmosphere, however, the spectral resolution was set
to about 130,000. The sets of synthetic spectra were then compared to the
observations to find the model with the best fitting lithium line to
estimate the abundance of lithium.  For the comparison, the synthetic
spectra were convolved with a rotational profile of $15\,$km/s to account
for the rotational broadening of the star (Kenyon \& Garcia 1986) and the
observed spectrum was converted to vacuum wavelengths. In addition, a
global blue-shift of $31\,$km/s was applied to the synthetic spectra to
match observed and computed features. We have also calculated spectra for
effective temperature and gravities close to the best fitting model to
establish the error in the estimated lithium abundance. For the purpose
of the analysis presented here, we used a fixed micro-turbulent velocity
$\xi=2\,$km/s.  In all cases, the quality of the fits was established
manually by inspection of the plots comparing the synthetic spectra to
the data.

The Li 6708\AA\ resonance line is well known to be very sensitive to
T$_{\rm eff}$. Thus we tried different temperatures between 3000 K, and
3600 K in our abundance analysis. We also tried three different values
for the gravity; $\log$~g=2.5, $\log$~g=3.0 and $\log$~g=3.5. The best
fitting model for T CrB [$T_{\rm eff}=3200\,$K, $\log(g)=2.5$] showed the
Li abundance to be between 0.5 and 0.7. We estimate uncertainies of
$\sim$200K and $\sim$0.5 in $T_{\rm eff}$ and $\log(g)$ respectively. As
shown in Figure 1 we obtained a good representation of the spectrum
without changing the abundances of the other elements (Al, Ca, Fe, Si)
present in the synthesised spectral region. Note that the Fe$\sc i$
6707.44\AA\ line is included in the synthetic spectrum. 

A similar analysis for the standard star HD~151203 shows that its
Li abundance is at least 0.4 dex below that of T CrB.
As one can see from Figure 1 the Li line is not well matched, suggesting
that there may be a line missing from the synthetic spectrum, accounting
for the missing absorption. However, this 
could well be that the CN line data, which is included in the model, not
being very accurate.

\section{Discussion}

Normal stars reach sufficiently high temperatures to destroy lithium in
their interiors. As a result, if there is significant convection of
material to the surface from regions hot enough to destroy lithium, its
abundance will decline with age. This effect is seen for cool stars,
whilst the surfaces of hotter stars, within which convection does not
occur, retain what is thought to be their initial lithium abundance.
Although F-stars break this monotonic relationship, it is possible to say
that main-sequence stars of $>$1.5$M_{\odot}$ do not show significant
depletion (Balachandran 1988). Once the stars leave the main sequence,
the surface abundance of the hotter stars depends crucially on the onset
of convection. This yields a possible explanation of the lithium we have
detected. The secondary star in T CrB was probably initially greater than
1.5$M_{\odot}$, and although the star has become a giant, convection or
dredge-up may not yet have begun. Just such an explanation is used for
the few normal giants which have almost solar abundance lithium, although
most giants are, as one would expect, lithium poor (Brown et al 1989).

Another explanation of the lithium abundance may be provided by
comparison with other late-type giants. Pallavicini, Randich and Giampapa
(1992) show that coronally active K-giants have relatively strong lithium
lines. Although sunspots also show strong lithium, starspots alone cannot
be the cause of the enhanced lithium line (Pallavicini et al 1993), and
it is thought to reflect a true abundance anomaly. The reason for this is
unclear, but ideas related to a lack of differential rotation as a
function of radius, would fit in with the observation that tidal locking
can also inhibit lithium depletion. However, it should be noted that
there is observational evidence for this is very limited.
Both mechanisms should be present in T CrB, helping to explain
the high lithium abundance, although it should be noted that the
latest spectral type studied by Pallavicini et al (1992) was K, not
M. 

A final mechanism which may enhance the surface abundance is material
placed there by the nova explosion. Unfortunately the production of
lithium in novae is controversial, with estimates for the abundance in
the ejecta varying from solar (Boffin et al. 1993) to several hundred
times solar (Starrfield et al 1978). The observations may of be some
help here, since the old nova GK Per shows no lithium enhancement
(Martin et al 1995). This would argue that novae do not significantly
affect the lithium abundance of their secondary stars, although one
should be cautious of extrapolating from one peculiar cataclysmic
variable (GK Per), to another (T CrB).

To summarise the above three paragraphs, there are three possible
mechanisms for the lithium we observe. It may be that the giant has
not yet become convective, it may be related to tidal locking and
stellar activity, or it may be due to the nova explosion. Note that we
cannot rule out any of these ideas.


\section{Conclusion}

We have obtained high-resolution echelle spectra of T CrB in the Li
6708\AA\ spectral region. Spectral synthesis fits gives the Li
adundance to be $\log$~N(Li)$\sim$0.6 , a factor of 4 below solar. The Li
abudance in field stars of the same spectral type as the secondary
star in T CrB is not detectable. We offer three possible explanations
for the enhancement of Li in T CrB. It is either due to the a delay
in the onset of convection in the M-giant, stellar activity on the
surface of the companion star or it may be due to the nova explosion.

\section*{Acknowledgments}

This work was based on
observations collected using the Bernard Lyot Telescope at the Pic du
Midi Observatory. This work was supported in part by NSF grant
AST-9720704, NASA ATP grant NAG 5-3018 and LTSA grant NAG 5-3619 to the
University of Georgia. Some of the calculations presented in this paper
were performed at the San Diego Supercomputer Center, with support from
the National Science Foundation, from the NERSC, and from the U.S. DoE.
TN was supported by a PPARC Advanced Fellowship.

\begin{figure}
\caption{NLTE synthesis fit to the Li 6708\AA\ region. The upper
panel shows a fit to the T CrB (summed) spectrum using T$_{\rm eff}$=3200
K, $\log$~g=2.5. Models for $\log$~N(Li)=0.4, 0.6 and 0.8 are shown. The
lower panel shows a fit to HD~151203 (M3$\sc iii$) using T$_{\rm
eff}$=3600 K, $\log$~g=2.5. Models for $\log$~N(Li)=0.0 and 0.2 are
shown.}

\end{figure}

\begin{figure*}
\rotate[l]{\epsfxsize=600pt \epsfbox[-100 800 800 800]{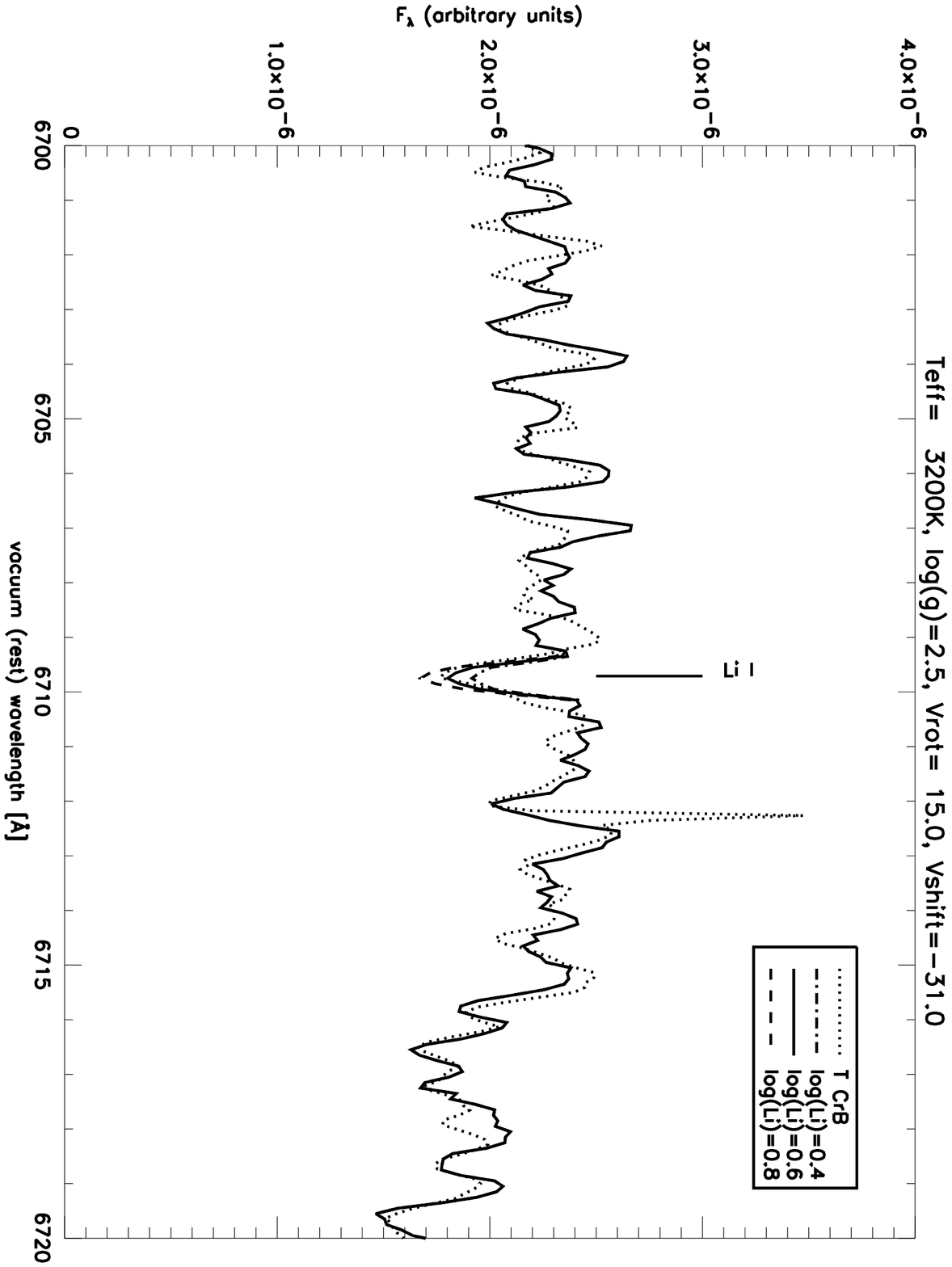}}
\end{figure*}

\begin{figure*}
\rotate[l]{\epsfxsize=600pt \epsfbox[-100 800 800 800]{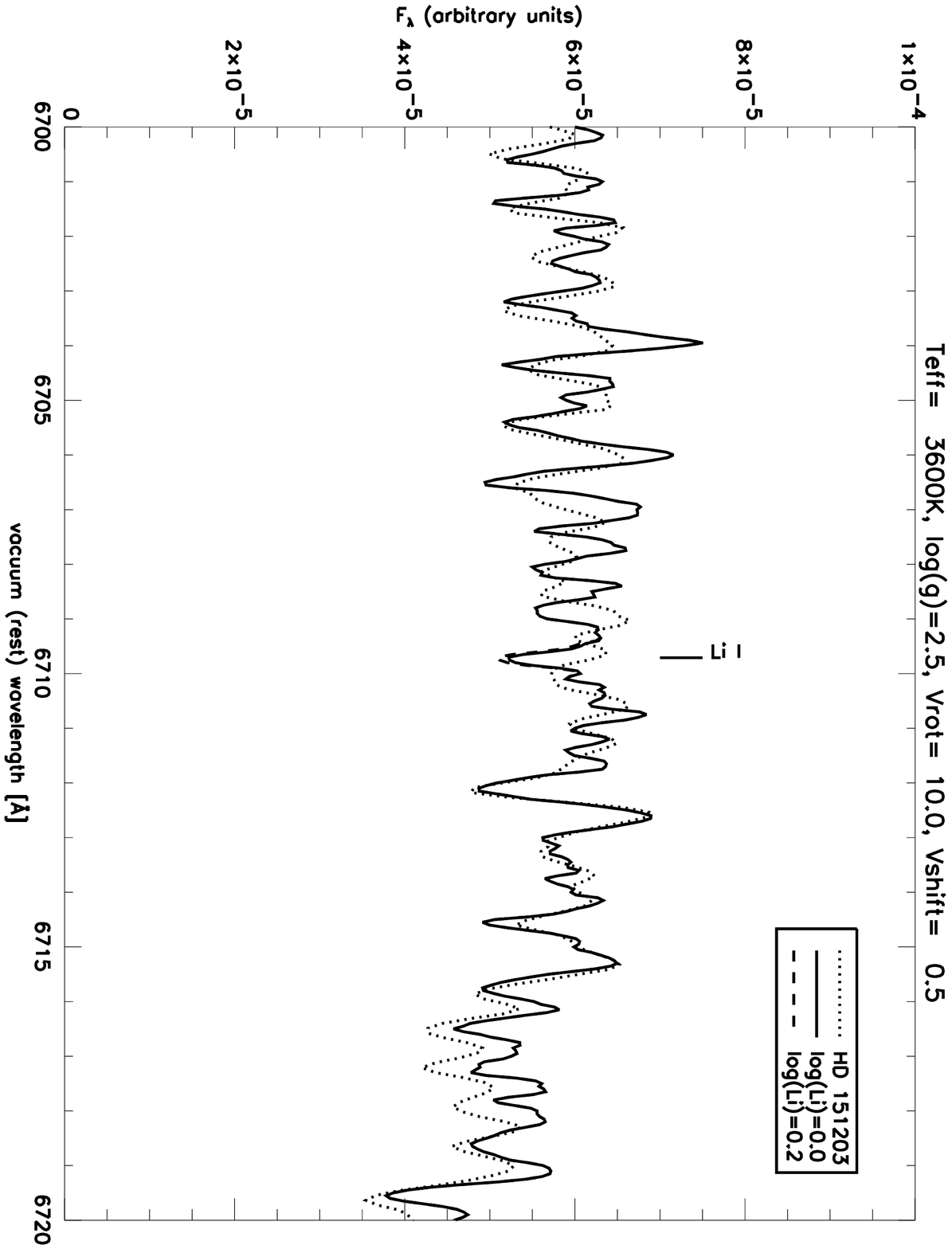}}
\end{figure*}


\begin{thebibliography}{}

\bibitem{} Balachandran S., 1988, PhD thesis, University of Texas at Austin

\bibitem{} Boffin H.M.J., Paulus G., Arnould M., Mowlavi N., 1993,
A\&A, 279, 173

\bibitem{} Brown J.A., Sneden C., Lambert D.L., Dutchover E., 1989,
ApJS, 71, 293

\bibitem{} Hauschildt, P.H., Allard, F., \& Baron, E. 1998, ApJ, in press

\bibitem{} Kenyon S.J., Garcia M.R., 1986, AJ, 91, 125

\bibitem{} Livio M., Truran J.W., 1992, ApJ, 389, 695

\bibitem{} Martin E.L., Casares J., Charles P.A., Rebolo, R., 1995, 
A\&A, 303, 785

\bibitem{} Pallavicini R., Randich S., Giampapa M.S., 1992,
A\&A, 235, 185

\bibitem{} Pallavicini R., Cutispoto G., Randich S., Gratton R., 1993, 
A\&A, 207, 145

\bibitem{} Shahbaz T., Somers M., Naylor T., 1997, 288, 1027

\bibitem{} Starrfield S., Truran J.W., Sparks W.M., Arnould M., 1978, 
ApJ, 222, 600

\bibitem{} Webbink R.F., 1976, Nat, 262, 271

\end{thebibliography}
\end{document}